\documentclass[12pt]{article}
\usepackage{color}
\usepackage{graphics}
\usepackage{rotating}
\usepackage{amsfonts}
\usepackage{epsfig}
\usepackage{amsmath}
\def\one{{\hbox{1\kern-.8mm l}}}

\newcommand{\beq}{\begin{equation}}
\newcommand{\eeq}{\end{equation}}
\newcommand{\be}{\begin{eqnarray}}
 \newcommand{\ee}{\end{eqnarray}}

 \newcommand{\half}{\frac{1}{2}}

\newcommand{\ov } {\over }
\newcommand{\p }{\partial }
\newcommand{\bp }{ {\bar\partial } }
\newcommand{\s }{\sigma }
 
\newcommand{\T }{ {\cal T } }

\def\td{\tilde }

\def\bt{ {\bar \tau} }
\newcommand{\field}[1]{\mbox{\textsf{#1}\hspace{-5.6pt}\textsf{#1}}}

\begin{document}
\null\vskip-24pt 
\hfill
SISSA-43/2004/EP 
\vskip-1pt
\hfill {\tt hep-th/}
\vskip0.2truecm
\begin{center}
\vskip 0.2truecm {\Large\bf 
Long Lived Large Type II Strings: decay within compactification
}

\vskip 2.5truecm

{\bf Diego Chialva $^{a,b}$
, Roberto Iengo $^{a,b}$ 
}

\bigskip
\medskip


{{${}^a$
\it International School for Advanced Studies (SISSA)\\
Via Beirut 2-4, I-34013 Trieste, Italy} 

\medskip

{${}^b$ INFN, Sezione di Trieste}


\bigskip

{\tt{chialva@sissa.it},
\tt{iengo@he.sissa.it}
}}

\end{center}
\begin{abstract}

Motivated also by recent revival of interest about metastable string
states (as cosmic strings or 
in accelerator physics), we study the decay, in presence of dimensional 
compactification, of a particular superstring state, which was proven
to be remarkably long-lived in the flat uncompactified 
scenario. We compute the decay rate by an exact numerical evaluation
of the imaginary part of the one-loop propagator. For large 
radii of compactification, the result tends to the fully
uncompactified one (lifetime 
$\mathcal{T}\equiv {\it const}g_s^{-2}M^5$), as expected, the string 
mainly decaying by massless radiation. For small
radii, the features of the decay 
(emitted states, initial mass dependence,....) change, depending on how
the string wraps on the compact dimensions.

\end{abstract}

\date{June 2004}

\vfill\eject

\tableofcontents
\listoftables
\listoffigures

\setcounter{section}{0}

\section{Introduction and Summary}

It is important to explore the genuine stringy features of SuperString
theory, in
particular the properties of its massive excited states, both for a
better understanding  
of an essential ingredient of the theory and also in view of possible 
phenomenological implications in cosmology, see for instance the
recent renewed interest in cosmic strings \cite{cosmpol,intcosmpol}, or even in
accelerator physics.

Those kind of investigations have a long story, even though rather
diluted in years
(\cite{Green,Mitchell,Dai,Okada,Sundborg,Wilkinson,Turok,Amati2,IK,Manes,IR1,
IR2,CIR}).
One of the relevant questions is whether  there could be large 
"macroscopic" metastable string states. In a previous paper
(\cite{CIR}) it was found an
affirmative answer to that question: by studying the decay properties of 
some Type II string states it was found that for a particular
configuration the lifetime
was increasing like the $\it {fifth}$ power of the mass. 
This result came from a detailed numerical investigation of the exact
quantum expression
of the decay rate and  
it was obtained within the basic layout of the theory, that is for the case
of $9+1$ dimensions, when all of the $9$ space dimensions were uncompactified.
The whole decay pattern had a simple interpretation 
which also suggests how compactification would affect its main features.

The scope of the present paper is to present a detailed study of the
decay properties
of the previously found long-lived TypeII Superstring state in the case
when some of 
the space dimensions are compactified. The results are in overall
agreement with the expected pattern. 

The required computations are standard, 
but the numerical evaluation is quite tough.
For that reason we limit our study to the case in which two of the 
$9$ dimensions are (toroidally) compactified. However the results are 
sufficiently clear
to show in general the main implications of the compactification on the decay 
properties of our long lived state and thus also clearly indicate what
would be the result
for a complete compactification.

In order to introduce and describe the present work, it is necessary to recall 
the main points of \cite{CIR}. It was introduced there a family of TypeII superstring
states,
which had a classical interpretation: the string of this family  lies
in general in $4$ space dimensions 
(that is necessary for satisfying the Virasoro constraints) and,
depending on a parameter, it describes two ellipsis in two orthogonal
planes. In one limiting case it reduces to 
a folded string, rotating in just one plane (classically like the
state of maximal angular momentum,
although quantum mechanically not precisely so); in the opposite
limiting case it describes 
two rotating circles, with opposite chiralities, in  two orthogonal
planes. The length of the string,
and therefore its mass, can be arbitrarily large.

The vertex operator corresponding to the exact superstring state was
derived and 
used to construct  the loop amplitude giving the  quantum correction
to its mass.
The imaginary part of this mass shift gives the decay rate: it is
expressed as a modular invariant integral of some combination of theta 
functions, whose imaginary
part is numerically evaluated using standard techniques.

The results show very clearly that while the folded string 
rather easily decays by classical breaking, and actually the numerical results 
reproduce with astonishing accuracy the classical pattern, the
circular rotating string does not break,  
rather it slowly decays by soft emission of massless 
 particles (gravitons, etc). In the present paper we will study this last case,
corresponding to the following classical  string configuration:
\beq \label{solution}
X^1+iX^2=Le^{i(\tau +\sigma)} ~~~~~~ X^3+iX^4=Le^{i(\tau -\sigma)}
\eeq
 
To be precise,  the breaking of the rotating circular string into two
massive fragments was found to be
not absolutely forbidden, but severely exponentially suppressed, 
and thus completely negligible, for large sizes of the parent state.  
This agrees with the picture that the string 
interactions are described by the overlap of the initial and final 
configurations.
By the classical world-sheet equations of motion 
this requires continuity of the string and its derivative in the
world-sheet time 
(see examples in \cite{IR2}).
It is seen that no possible breaking of the string eq.(1) can meet the
classical requirements, 
neither into closed nor into open fragments, neither with free
(Neumann) nor with fixed 
end-points (Dirichlet) b.cs.
\footnote{In brief, open string coordinates with either of the
  b.cs. will be of the form
$X(\tau ,\sigma )=f(\tau +\sigma )\pm f(\tau -\sigma )$ which does not
  matches with eq.(1).}
 The breaking can occur via quantum fluctuations,
but it is suppressed if they are required to be large. 

A closed string could also be absorbed by a D-brane leaving it in an
excited state,
see \cite{hashkleb}; the matrix element of the quantum state corresponding to
the string eq.(1)
with a D-brane, however, is 
either zero or it is for large $L$ in general exponentially suppressed,
except for some particular arrangements of angles of the string planes
with respect to
the tangent and orthogonal directions of the brane.
Also, its coupling to a homogeneously decaying brane, see
\cite{lamblimal}, is further 
suppressed by an exponential in the energy.
In presence of branes there could be also other decay modes, discussed
in \cite{cosmpol},
which would be allowed for some particular arrangements of branes and fluxes.
We do not know if those processes can be described by classical equations 
and it would be interesting to see an actual computation of those decay rates 
for various string states. 
Here  we assume a background configuration in which those coupling 
to the branes or those modes do not occur or are suppressed. 

The total rate of the dominant decay channel of the circular rotating string, 
that is the soft emission of massless modes, was found to be
proportional to $M^{-5}$ 
where $M$ is the mass of the decaying state. 
This is in agreement with a sample computation using the operatorial
formalism which gives a rate proportional to $M^{-d+4}$ where $d$  is
the number of large space dimensions,
$d=9$ in \cite{CIR}.  The same estimate can be obtained by a
semi-classical reasoning:
the rate of massless production is $\omega^{d-2}/M^2$ (a phasespace
factor, $\omega$ being
the massless particle energy) times the modulus square of the strength
of the source,
for instance the energy-momentum tensor for graviton emission, which gives 
a factor $|L^2|^2\sim M^4$ , the length of the state being
proportional to its mass $L=\alpha^{'} M$.
Since $\omega = {2m\ov \alpha^{'} M}$, and $m$ 
is found to be bounded, it results the above estimate.

Thus we expect a decay rate proportional to $M^{-d_{eff}+4}=M^{-3}$ (here $d=d_{eff}=7$)
for two compactified dimensions,
when the size of the compact dimensions is small with respect to the
string size and the
string lies the uncompactified space. If instead the dimensions' size
is large as compared to the string size 
we expect to recover the fully uncompactified result.  

If however the string lies in part  or $\it{in ~ toto}$ in the compact 
dimensions  and it winds around them,  we expect a completely
different result , because in this case the breaking is possible. To
be precise, we expect the 
breaking to easily occur when the string winds around one dimension
with one chirality
and around the other dimension with the opposite chirality, in order
to respect the Virasoro
constraints.

 The results are relevant in particular for a possible physical
scenario in which 
$4$ of the $6$ compactified dimensions are of the order of $\sqrt{\alpha^{'}}$
but $2$ of them are much larger \cite{Antoniadis}. 
The rotating circular string could for instance lie in the space
spanned by two uncompact
and two large compact dimensions, and it could  be as large as the
size of the latter ones 
or even more, depending on how it winds and how easily it can break
into winding modes.
This string would slowly decay by soft massless radiation, with a
total lifetime of the 
order of $M^2$ (a factor $M$ from the inverse rate times  another
factor $L\sim M$,
from the time necessary to substantially lowering the string size). 
A detailed study of this picture and its possible implications will be
discussed elsewhere. 

 This paper is organized as follows. In Section \ref{imaginarypart} 
we outline the
main points of the computation,
leaving the full explicit description to the Appendix
\ref{appendiximm}. 
We have here 
considered, as said above, two compact
and seven noncompact dimensions, and we have made the computation for
three cases:
1) the subspace in which the string lies is totally uncompact, 2) the
subspace in which the string lies 
has one compact dimension, 3) the subspace in which the string lies
has two compact dimensions.
In Section \ref{dataanalysis} we present the numerical results and we discuss
them. In Section \ref{conclusion} we draw
the conclusions. In Appendix \ref{appendixtables} we report a relevant
selection of data. 
    
\section{The imaginary part of the mass-shift.} \label{imaginarypart}
We will consider a space-time configuration having 7+1 extended and 2
compactified coordinates.

From now on we will choose
 \beq
 \alpha'=4
 \eeq
and define the complex coordinates
 \beq \label{complexcoord}
 Z_1=\frac{X_1+iX_2}{\sqrt{2}},\quad Z_2=\frac{X_3+iX_4}{\sqrt{2}}.
 \eeq
and analogous for the fermionic partners.

We are going here to show the expressions for the imaginary part of
the one-loop mass shift for the quantum state (we follow the notation
of \cite{CIR})
 \beq
 |\Phi \rangle = \mathcal{N}\ |\phi^R  \rangle  |\phi^L  \rangle
 \label{complete}
 \eeq
with
 \beq
 |\phi^R\rangle=
 (\psi^{\rm z_1}_{-\half})^\dag
 (b_{1}^\dag)^{N}|0, p\rangle \ ,
 \label{estado}
 \eeq
and $\mathcal{N}$ a normalization constant. $b_{1}^\dag$ is the level
one right creation operator in the expansion of $Z_1$ and 
$\psi^{\rm z_1}_{-\half}$ is the level one half creation operator for
the fermionic coordinate $\psi^{\rm z_1}$. 
We have written  only the right-moving part: 
the left-moving is obtained by substituting $b_1^\dag$  with
$\td c_1^\dag$ (the level one creation operator in the expansion of
$Z_2$) and $\psi^{\rm z_1}_{-\half}$ with $\td \psi^{\rm z_2}_{-\half}$. 

The normalization constant in front of the state can be easily
computed by requiring $\langle \Phi |\Phi \rangle =1$, giving:
\beq
\mathcal{N} = \frac{1}{N!}\ .
\label{norm}
\eeq

We are going to compute the imaginary part of the world-sheet torus amplitude 
 \beq \label{vertexamplitude}
 \Delta M^2=\int \frac{d^2\tau}{\tau_2}\int d^2z\langle \bar
 V(z)V(0)\rangle
 \eeq
where $V(z)$ is the vertex operator for the state 
(\ref{complete}) (derived in \cite{CIR}):
 \beq \label{vertexoperator}
 V=\frac{1}{2^{N-1}}\frac{N^2}{N!}\psi_{z_1}\p\psi_{z_1}(\p
 Z_1)^{N-1}\td\psi_{z_2}\bp\td\psi_{z_2}(\bp Z_2)^{N-1}e^{ipX}
 \eeq
whose mass (in units $\alpha'=4$) is $M^2=N$. 

 We must now distinguish the cases where the string lies on the
compactified coordinates from the one where the string propagates
only in the flat extended directions.

\subsection{String lying on extended dimensions only.}

Let us suppose that two space-time coordinates, different from
$X^1,X^2,X^3,X^4$, are compactified on two circles of radii $R_1,
R_2$. The computation of formula (\ref{vertexamplitude}) is
straightforward, since the propagator for
the string is the same as in the uncompactified scenario; for the
details look at appendix (\ref{appendixsu0}).

The amplitude is: 
 \be \label{amplitudeextendedstring}
 && \Delta M^2_{k,n}=c' \ g_s^2\ \int {d^2\tau\ov\tau_2^4}
  \int d^2z \ e^{-4N{\pi y^2\ov\tau_2}}
  \left| {\theta_1 (z|\tau)\ov\theta_1^{'}(0|\tau)}\right| ^{4N}
  \big( {\pi\ov\tau_2} \big) ^{2N-2} \nonumber\\
 &&\times \sum_{m_1,m_2} \big( {\pi\ov\tau_2} \big) ^{-m_1-m_2}
  qq(N;m_1,m_2) \big( \p^2 \log(\theta_1(z|\tau ) \big)^{m_1}
  \big(\bar\p^2\log(\theta_1(z|\tau ) \big)^{m_2}\nonumber \\
 && \times\prod_{i=1}^2\frac{1}{R_i}
  e^{\sum_{n_i,w_i}2i\pi\tau\left({n_i\ov R_i}+{w_iR_i\ov 4}\right)^2-2i\pi\bar\tau\left({n_i\ov R_i}-{w_iR_i\ov 4}\right)^2}
 \ee
with $qq(N;m_1,m_2)=\frac{N^2((N-1)!)^2}{m_1!m_2!(N-1-m_1)!(N-1-m_2)!}$
and $c'$ a numerical constant.

\vspace{0.5cm}

It is convenient to expand:
 \beq \label{pqexpansion}
 (e^{i2\pi z})^N \left(
 {2\pi\theta_1 (z|\tau )\ov\theta_1^{'}(0|\tau)}\right) ^{2N} \big(
 {1\ov 4\pi^2}\p^2 \log(\theta_1(z|\tau ) \big) ^{m_1} =
 \sum_{p,q}\gamma (N,m_1;p,q)\ e^{i2p\pi\tau}e^{i2(q-p)\pi z}
 \eeq
and similarly for the antiholomorphic part with coefficient $\gamma
(N,m_2;\tilde p,\tilde q)$ (that we obtain through computer calculation).

We have to compute the imaginary part of the integral:
 \be
 H &\equiv &\int \frac{d^2\tau}{\mbox{\small{$\tau_2^{2+2N-m_1-m_2}$}}}
              e^{2i\pi\tau_1\left(p-\td p+\sum_{i=1}^2w_in_i\right)-
               2\pi\tau_2\left(p+\td p+\sum_{i=1}^2\left({2n_i^2\ov R_i^2}+
               {w_i^2R_i^2\ov 8}\right)\right)}
              \nonumber \\
   &\times &\int d^2z\, e^{2i\pi x(q-p-\td q+\td p)-2\pi y(q-p+\td q-\td p-2N)-4N\pi {y^2\ov \tau_2}} 
 \ee

Integration over $\tau_1$ e $x$ leads to:
 \be
 \begin{cases}
 p-\td p & =  -\sum_{i=1}^2w_in_i \\
 q-\td q & =  p-\td p  \\
 \end{cases}
 \ee

Comparing this expression with the Schwinger parametrization of the
one-loop two point amplitude $\langle \Phi\Phi\rangle$ of a field theory
with coupling $\Phi \to\phi_1\phi_2$, where
$\Phi$ has mass $M$ and $\phi_1,\,\phi_2$ have masses
$M_1,\, M_2$ (see \cite{IR1, CIR}), we find that
 \be \label{masse12}
 \begin{cases} 
 M_1^2& = {q+\td q\ov 2}+\sum_{i=1}^2\left({n_i\ov R_i}\right)^2+\left({w_iR_i
 \ov 4}\right)^2=q+\sum_{i=1}^2\left({n_i\ov R_i}+{w_iR_i\ov 4}\right)^2 \\
 M_2^2& = {p+\td p\ov 2}+\sum_{i=1}^2\left({n_i\ov R_i}\right)^2+\left({w_iR_i
 \ov 4}\right)^2=p+\sum_{i=1}^2\left({n_i\ov R_i}+{w_iR_i\ov 4}\right)^2 \\
 \end{cases}
 \ee

The imaginary part of the amplitude is therefore obtained by a
 standard formula (see \cite{Okada,IR1}), and we get 
($c''$ is a numerical constant):
 \be \label{imaginarypartsu0}
  Im(\Delta M^2) & = &
                {c^{''}g^2_s\ov \sqrt N}{4\ov \pi R_1R_2}
                 \sum_{p,q,m_1,m_2,w_1,n_1,w_2,n_2}4^{-(2N-m_1-m_2)}  \\
              & \times & \gamma(N,m_1,p,q)\gamma(N,m_2,p+\sum_{i=1}^2w_in_i,q+\sum_{i=1}^2w_in_i) \nonumber \\
              & \times & qq(N;m_1,m_2)
                 {\left[N\hat\omega-4\sum_{i=1}^2\left({n_i\ov R_i}+{w_iR_i\ov 4}\right)^2\right]^{2N-m_1-m_2+1/2}
                 \ov \Gamma(2N-m_1-m_2+3/2)} \nonumber
 \ee
where 
$\hat\omega =1-2({p\ov N}+{q\ov N})+({p\ov N}-{q\ov N})^2.$

The amplitude is normalized 
such that in the
limit $R_1,\,R_2 \rightarrow +\infty$ the result reproduces the
one in the flat extended ten dimensions scenario (\cite{CIR}).

\subsection{String lying on one compactified dimension (right-moving modes).}

Suppose now that the coordinate $X^1$ is compactified on a circle of radius
$R_1$ (only right-moving modes can wind) and another coordinate,
different from $X^2,X^3,X^4$, on a circle of radius $R_2$. In this
case the propagator of the string is modified. 
By the derivation presented in appendix (\ref{appendixsu1}), we find
an imaginary part for the amplitude (\ref{vertexamplitude}):

 \be \label{imaginarypartsu1}
 Im(\Delta M^2)&=&{c^{''}g^2_s\ov \sqrt N}{4\ov \pi
                   R_1R_2}\sum_{r=0}^{N-1}\sum_{m_1=0}^{N-1}\sum_{m_2=0}^{N-1}\sum_{s=r}^{N-1-m_1}
                   \sum_{p,q,w_1,n_1,w_2,n_2}4^{-(2N-m_1-m_2)}  \\
 & \times &\gamma(N,m_1,p,q)\cdot\gamma(N,m_2,p+\mbox{$\sum_{i=1}^2$}w_in_i,q+\mbox{$\sum_{i=1}^2$}w_in_i) \nonumber \\
 & \times & qq(N;m_1,m_2,s)\cdot C(s,r) \nonumber \\
 & \times & \left({2\ov R_1}n_1+{R_1\ov 2}w_1\right)^{2r}{\left[N\hat\omega-4\sum_{i=1}^2\left({n_i\ov R_i}+{w_iR_i\ov 4}\right)^2\right]
                 \ov \Gamma(2N-m_1-m_2-r+3/2)}^{2N-m_1-m_2-r+1/2} \nonumber
 \ee
where
 \be
 qq(N;m_1,m_2,s)& = &{N^2((N-1)!)^2\ov
                     (s!)^2(N-1-m_1-s)!(N-1-m_2)!m_1!m_2!} \\
 C(s,r)& = &\sum_{l=r}^s c_s^l{\Gamma(3/2)\ov \Gamma(3/2-l+r)}(-1)^{s+r}
            \mbox{\small{$\left(\begin{array}{c} l\\r \end{array}\right)$}}
 \ee
($c_s^l$ are defined in (\ref{cslcoefficients}) in appendix \ref{appendixsu1}).
Also here, comparison with the relevant field theory amplitude
gives us equation (\ref{masse12}).

\subsection{String lying on two compactified dimension ( left- and right-moving
 modes).}
Suppose now that the coordinates $X^1,X^3$ are compactified on two
circles of radii $R_1,R_2$. The computation shown in appendix 
(\ref{appendixsu2}) leads to the following imaginary part for
amplitude (\ref{vertexamplitude}):
 
 \be \label{imaginarypartsu2}
 Im(\Delta M^2)&=&{c^{''}g^2_s\ov \sqrt N}{4^{1-2N}\ov \pi
                   R_1R_2}\sum_{r_1=0}^{N-1}\sum_{r_2=0}^{N-1}\sum_{m_1=0}^{N-1}\sum_{m_2=0}^{N-1}\sum_{s_1=r_1}^{N-1-m_1}\sum_{s_2=r_2}^{N-1-m_2}
                   \sum_{p,q,w_{1,2},n_{1,2}}4^{(m_1+m_2)}  \nonumber \\
 & \times &\gamma(N,m_1,p,q)\cdot\gamma(N,m_2,p+\mbox{$\sum_{i=1}^2$}w_in_i,q+\mbox{$\sum_{i=1}^2$}w_in_i) \nonumber \\
 & \times & qq(N;m_1,m_2,s_1,s_2)\cdot C_1(s_1,r_1)\cdot C_2(s_2,r_2) \nonumber \\
 & \times &\left({2\ov R_1}n_1+{R_1\ov 2}w_1\right)^{2r_1}\left({2\ov
 R_2}n_2-{R_2\ov 2}w_2\right)^{2r_2} \nonumber \\
 & \times & {\left[N\hat\omega-4\sum_{i=1}^2\left({n_i\ov R_i}+{w_iR_i\ov 4}\right)^2\right]
                 \ov \Gamma(2N-m_1-m_2-r_1-r_2+3/2)}^{2N-m_1-m_2-r_1-r_2+1/2} 
 \ee
and this time
 \be
 qq(N;m_1,m_2,s_1,s_2)& = &\mbox{\normalsize{${N^2((N-1)!)^2\ov
                     (s_1!)^2(s_2!)^2(N-1-m_1-s_1)!(N-1-m_2-s_2)!m_1!m_2!}$}} \\
 C_i(s_i,r_i)& = &\sum_{l_i=r_i}^{s_i} c_{s_i}^{l_i}{\Gamma(3/2)\ov \Gamma(3/2-l_i+r_i)}(-1)^{s_i+r_i}
            \mbox{\small{$\left(\begin{array}{c} l_i\\r_i \end{array}\right)$}}.
 \ee
Again, through comparison, we get equation (\ref{masse12}).

\section{Data analysis.} \label{dataanalysis}

In this section we report the analysis of the data relative to the quantum
decay of the string (conventions: $ c^{''}= 32 (2\pi)^3$ (see
\cite{Okada}), $\alpha'=4$). 
We have
obtained these data through an exact computation performed using a
program written in Fortran90. The only approximation
is due to the precision of number representation used by the compiler.

We have calculated the imaginary part of the amplitude (\ref{vertexamplitude})
and from it the decay rate ($M$ is the mass of the state):
 \beq
 \mathcal{R}={\mathrm{Im}(\Delta M^2)\ov 2M}
 \eeq
and the lifetime:
 \beq
 \mathcal{T}={1\ov \mathcal{R}}.
 \eeq

The natural dimension-full scales of the decay are the radii of the
compactified dimensions and the length of the string, that is the
mass of the quantum state. We can expect that if the radii
of compactification are (much) larger then the length of the
string, the fact that those dimension are compact makes no
difference for the quantum process of decay (the string cannot
``see'' the compactification), instead things should change if the
string length is larger than the radii. Especially, if the string
can wrap on those directions, we expect a more rapid and intense
decay (through winding modes), since even classically different
points of the string can get in contact.

 Therefore we can formulate a
 mass-dependence for the lifetime, starting from the result obtained in ten 
flat dimensions in (\cite{CIR}) for our same state.
 The result found there was
 \beq
 \mathcal{T}=cg_s^{-2}M^5,
 \eeq
where $c$ is a proportionality constant.

Now call $R$ the radii of compactification (not necessarily all
identical: we will just for this moment set $R_i=R$, where $i$ runs on
the compactified dimensions).
The expectation we have about the lifetime is
 \beq \label{lifetimeexpected}
 \mathcal{T}=cg_s^{-2}M^{5-d_c}R^{d_c},
 \eeq
where $d_c$ are the compactified dimensions with
 \beq
 R\ll M.
 \eeq
This should be the right 
functional dependence when decays through winding (and
Kaluza-Klein at small radii of compactification) are
suppressed.

The dependence on $R$ of the lifetime in this case is suggested by formulas
(\ref{imaginarypartsu0},\,\ref{imaginarypartsu1},\,\ref{imaginarypartsu2}), 
that
show a common factor $R^{-d_c}$, with $d_c$ the number of
compactified dimensions (in the formulas $d_c=2$, since here there are
only two compactified dimensions).

As far as the dependence on $M$, 
the computation in \cite{CIR} 
of the decay of
the string in a graviton plus a massive state 
shows that the phase space factor should
be dimensionally reduced by the compactification at small radii, 
see the discussion in the introduction, resulting in formula (\ref{lifetimeexpected})

We have made two different analysis of our data.

To study
the radii dependence of the lifetime we have fitted the data assuming a
dependence (at $M$ fixed)
 \beq \label{fitradiusdependence}
 \mathcal{T}=cR^{\alpha}
 \eeq
and divided them in two sets, according whether $R\ll M$ (small
radii) or $R\gg M$ (large radii).

To study the mass dependence, we have fitted the data assuming a mass
power law (at $R$ fixed)
 \beq \label{fitmassdependence}
 \mathcal{T}=cM^{2*\beta}
 \eeq
and again diving them in two sets: one having $R\ll M$, the other with
$R\gg M$.

We also remark that for high masses the data collection 
requires long time-machine and processors' power, therefore in
some cases we have limited our computations to certain range of
masses only.

\subsection{General results for the decay rates.}\label{generalresults}

We have analyzed the decay of the string in three particular
configuration in the background $\field{R}^{1,7}\times T^2$:
 \begin{itemize}
 \item[-]0: string lying on extended dimensions only;
 \item[-]I: string lying on three extended and one compact dimensions; 
 \item[-]II: string lying on two extended and two compact dimensions,
 with opposite chirality.
 \end{itemize}

The general results for the decay are summarized in what follows. We
recall formula (\ref{masse12}) to understand the role of winding and 
Kaluza-Klein modes in contributing to the mass of the final decay products. 
 
\paragraph{Configuration 0}: the favoured decay channel is 
through one massless  and one massive decay product. More in
detail: for small radii of compactification
winding and Kaluza-Klein modes contribution is negligible. For large
radii, windings again do not contribute and Kaluza-Klein modes play
the role of components of the momenta and we recover the result
holding for uncompactified dimensions. For any radii, non-zero
oscillatory excitations in both final states represent an absolutely negligible
fraction of the total.

\paragraph{Configuration I}: for small radii of compactification
winding modes gives a small contribution (sizable only at low total
energy). The decay is entirely given by the
channels where the final states are
\textit{massless+massive} and \textit{Kaluza-Klein+massive}. For very
small radii the first one utterly predominates. For large
radii, the situation is the same as for configuration 0.

\paragraph{Configuration II}: for small radii of compactification the
decay through winding modes is the dominant one. Again, for larger
radii, we see the same behaviour for configuration 0. 

\vspace{0.5cm}

In the following sections we will study in more details the string
decays.

\subsection{String lying on extended dimensions only, two compactified dimensions.}
Throughout all this section, we will consider the case in which
the string lies in a background $\field{R}^{1,7}\times S^1\times
S^1$ but on extended dimensions only. We have calculated data for
various values of $R_1,\, R_2$, the two radii of compactification,
but for shortness, we will
discuss the results for the cases $R_1=R_2=R$.

For the string in this configuration we expect that there are no
significant contributions from final states with winding modes.
Indeed, as it is seen in table (\ref{rateres01}), case $dw=0$,
channels with no final windings
utterly predominates the total decay rate. Furthermore, the prevailing
decay appears to be the emission of a massless plus a massive
final states: as we said, the production of Kaluza-Klein modes at
small radii is suppressed, while at large radii they play the role of
components of the momenta in the large dimensions. The contribution of
decays in  massive states with oscillator number different from zero
in both the final decay products 
is only a negligible fraction of the
total, of order $10^{-5}$ for the lowest mass and radii of
compactification, down to $10^{-13}$ for the largest mass and radii.

\begin{figure}
\centerline{\epsfig{figure=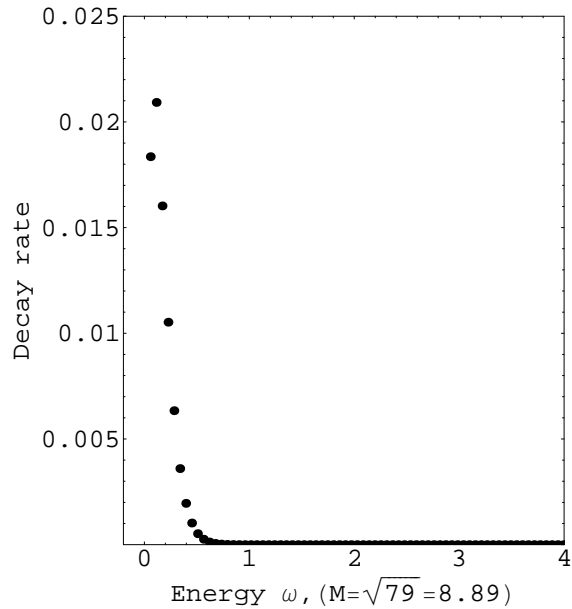}}
\caption[Emission spectrum for $R=2$.]{\footnotesize
Emission spectrum  for $R=2$ of the string lying 
on extended dimensions only.}\label{spectrsu0}
\end{figure}

In the uncompactified case ({\it i.e.} $R\to\infty$) the spectrum of 
the emitted massless particles shows a scale invariance:
when plotted in $m$ (with {\it energy}$=\omega={2m\ov \alpha' M}$) and
normalized to the same peak value, the spectra are almost identical
varying $M$, and
coinciding for higher masses.
We have verified that the same scaling happens also for small radii ($R<<M$).
In figure \ref{spectrsu0} we show the spectrum 
for $R=2$ and $M=\sqrt{79}=8.89$.
The dominant emission is a massless state
with a very low energy ({\it i.e.} $m$ limited for $M$ large)
plus another state of nearly the same mass of the initial one 
($M_1^2=M^2-{4\ov \alpha'}m$).
  
\begin{figure}
\vskip -0.5cm \hskip -1cm
\centerline{\epsfig{figure=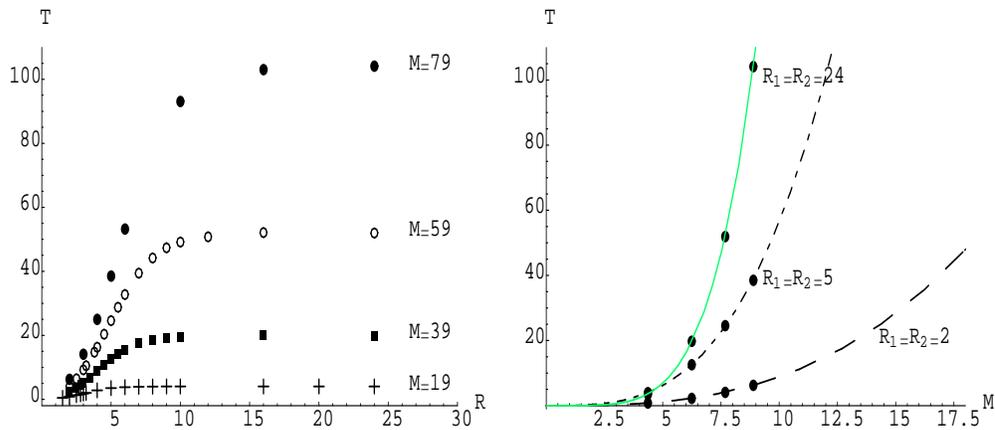,height=6truecm, width=14truecm}}
\caption[$R$ and $M$ dependence for the lifetime
(string lying on extended dimensions).]{\footnotesize
Case 0: $R$ and $M$ dependence for the lifetime.
The lines are fits to the data.}\label{Tsu0}
\end{figure}

\vspace{-0.5cm}
\subsubsection{$R$ dependence.}

We have performed the analysis considering values for the masses
of:
 \beq
 M^2=19,39,59,79.
 \eeq
These values have been chosen in accordance with the previous work
\cite{CIR}.

For each of them we have data for a range of values for the radii
$R$ starting from the string length $R=\sqrt{\alpha'}=2$, up to
values sufficiently higher than the threshold $M$.

We fit the dependence of the lifetime on the radii of
compactification as in (\ref{fitradiusdependence}). 
The result shows agreement with the expected power law 
(\ref{lifetimeexpected}), as can
be seen in table (\ref{Rdependencenonarr}). 
 \begin{table}[htbp]
 \centering
 \begin{tabular}{||c|l l||}
 Masses & $\alpha $ &    \\
 \hline
 $\sqrt{19}=4.36$ & 1.92231  & ($2\leq R\leq 3.2$)   \\
 $\sqrt{39}=6.24$ & 1.83862  & ($2\leq R\leq 5$)     \\
 $\sqrt{59}=7.68$ & 1.9104   & ($2\leq R\leq 6$)  \\
 $\sqrt{79}=8.89$ & 1.95665  & ($2\leq R\leq 6$)  \\
 \end{tabular}
 \parbox{5in}{\caption[$R$ dependence for the different masses
 (string lying on extended dimensions).]{
 $R$ dependence for the different masses: $\T \sim R^{\alpha}$. 
 See the discussion in the text for the uncertainty on $\alpha$.
 } \label{Rdependencenonarr}}
 \end{table}

The values found for $\alpha$ for radii $R<M$ approximate the expected
 \beq
 \alpha=d_c=2
 \eeq
within the 10\% and the agreement is better for the highest mass. Notably
(\ref{lifetimeexpected}) is already correct with a
good approximation for radii up to values close to the threshold.

Since the dependence on $R$ for small radii of compactification
just reproduces the factor $R^{-2}$ which
is in front of the amplitude (\ref{imaginarypartsu0}),
this is another confirmation that the contribution to the decay rate 
of the Kaluza-Klein 
modes different from zero is suppressed as well as the one of
windings.   

For radii sufficiently larger than $M$, the dependence on $R$
disappears as expected, with values of $\alpha$ in the range
 \beq
 -0.007 <\alpha < 0.030.
 \eeq
Also for large radii, the
 main contribution to the decay rate is by the channels with no winding modes
and no oscillator modes in both the final states. Therefore, since Kaluza-Klein modes
for large radii
play the role of components of momenta, we conclude that again the favoured 
decay is
through the emission of a massless and a massive mode.  

\subsubsection{$M$ dependence.}
 Since the string lies on extended dimensions only, the contribution
 to the decay from channels in which there are winding modes is
 expected to be suppressed. Therefore there are strong motivations
 for the mass dependence (\ref{lifetimeexpected}) also
 for small radii.

 Indeed the expectation is satisfied up to a reasonably good
 approximation, as can be
 seen from table (\ref{Mdependencenonarr}).
 \begin{table}[htbp]
 \centering
 \begin{tabular}{||c|l|c||}
 \multicolumn{3}{|c|}{Range of masses: $4.36<M<8.89$}\\
 \hline
 Radii & $\beta $ & Expected value   \\
 \hline
 2 & 1.42505 & 1.5   \\
 3 & 1.44265 & 1.5     \\
 4 & 1.53377 & $1.5^*$  \\
 5 & 1.5895  & $\mbox{}^*$  \\
 10 & 2.2036 & $\mbox{}^*$  \\
 16 & 2.27814 & 2.5  \\
 24 & 2.28217& 2.5  \\
 \end{tabular}
 \parbox{5in}{\caption[$M$ dependence for different radii 
 (string lying on extended dimensions).]{
 $M$ dependence for different radii: $\T \sim M^{2*\beta}$. 
 (*) indicates cases for which there is no clear expectation of the
 value of $\beta$ since $R\sim M$ for some of the masses considered.
 }\label{Mdependencenonarr}} 
 \end{table}

 The differences with the expected values can be accounted for in many ways: 
 from the approximation due to the precision of the number
 representation in the computer program used, but especially from the
 considered values of the masses and radii, that are probably not
 enough to explore the
 asymptotic limit.

\subsection{String lying on one compact and three extended dimensions.}

We consider here the configuration in which the string lies on one
of the compact dimensions in the space-time background
$\field{R}^{1,7}\times S^1\times S^1$.
The masses taken in account are $M^2=19,39,59$, for various
values of the compactification radii.

Semi-classical considerations suggest
that also in this case the decay rates  through winding modes 
should be suppressed, due to the fact that the Virasoro constraint are
less easily satisfied because of the left-right asymmetry of the
configuration.
Indeed, we have
verified that the decays in winding modes give just a very small
contribution, sizable only at low masses and radii.
We see also that the favoured decay is through
the emission of one Kaluza-Klein or massless and one massive states 
(see table (\ref{rateres01}), case $dw=1$) and
that this last case (massless plus massive final states) prevails for
very small radii ($R<<M$). Furthermore, the contribution of
decays in  massive states with oscillator number different from zero
in both the final states
is only a negligible fraction of the
total, of order $10^{-3}$ for the lowest mass and radii of
compactification, down to $10^{-11}$ for the largest mass and radii considered.

\begin{figure}[h!]
\label{Tsu1} 
\vskip -0.5cm \hskip -1cm
\centerline{\epsfig{figure=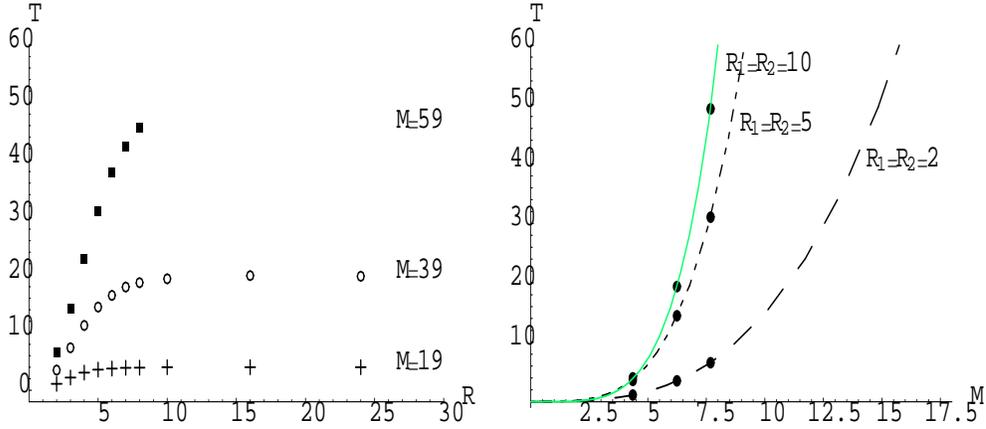,height=6truecm, width=14truecm}}
\caption[$R$ and $M$ dependence for the lifetime 
(string lying on one compact dimension).]{\footnotesize
Case I: $R$ and $M$ dependence for the
lifetime. The lines are fits to the data.}
\end{figure}

We show in figure \ref{Tsu1} the $R$ and $M$ dependence of the
lifetime.

\vspace{0.5cm}

\subsubsection{$R$ dependence.}
For radii $R<M$ the dependence (\ref{lifetimeexpected}) is
verified with increasingly good approximation at higher masses
(around 26.5\% of error for small masses, 19.5\% for higher ones, see
table \ref{Rdependencesu1}), 
suggesting that also the Kaluza-Klein contribution to the decay is
quite suppressed.
 \begin{table}[htbp]
 \centering
 \begin{tabular}{||c|l l||}
 Masses & $\alpha $ &    \\
 \hline
 $\sqrt{19}=4.36$ & 1.47213  & ($2\leq R\leq 4$)   \\
 $\sqrt{39}=6.24$ & 1.55633  & ($2\leq R\leq 5$)     \\
 $\sqrt{59}=7.68$ & 1.60627   & ($2\leq R\leq 6$)  \\
 \end{tabular}
 \parbox{5in}{\caption[$R$ dependence for the different masses 
 (string lying on one compact dimension).]{
 $R$ dependence for the different masses: $\T \sim R^{\alpha}$. 
 See the discussion in the text for the uncertainty on $\alpha$.
 }\label{Rdependencesu1}} 
 \end{table}

For radii $R>M$ and masses $M^2=19,\, 39$ the fit 
(\ref{fitradiusdependence}) gives values
of $\alpha$ in the range
 \beq
 0.03 <\alpha < 0.05,
 \eeq
that is the lifetime is $R$ independent 
(while for mass $M^2=59$ the value found is
 $
 \alpha=0.33887,
 $
that can be explained noticing that the values of the radii considered
are not so larger than the threshold).

\subsubsection{$M$ dependence.}
The broad qualitative picture for the mass dependence of the lifetime
is similar to the expected 
(\ref{lifetimeexpected}) but the transition between $\beta=1.5$,
(expected for small radii) and $\beta=2.5$ (expected for large radii)
 is less sharp
than in the case of the string lying on extended dimensions for the
data $R<M$ (see table (\ref{Mdependencesu1})).
 
\begin{table}[htbp]
 \centering
 \begin{tabular}{||c|l|c||}
 \multicolumn{3}{|c|}{Range of masses: $4.36<M<7.68$}\\
 \hline
 Radii & $\beta $ & Expected value   \\
 \hline
 2 & 1.56233 & 1.5   \\
 3 & 1.64233 & 1.5     \\
 4 & 1.774497 & $\mbox{}^*$  \\
 5 & 1.90883  & $\mbox{}^*$  \\
 10 & 2.212286 & $2.5$  \\
 \end{tabular}
 \parbox{5in}{\caption[$M$ dependence for different radii 
 (string lying on one compact dimension).]{
 $M$ dependence for different radii: $\T \sim M^{2*\beta}$. 
 (*) indicates cases for which there is no clear expectation of the
 value of $\beta$ since $R\sim M$ for some of the masses considered.
 }\label{Mdependencesu1}}  
 \end{table}

This is due to the increasing contribution for higher radii
by the channels of decay
into Kaluza-Klein modes (plus a massive final state).

The trend towards the asymptotic pattern is clear, even though our
analysis could only consider limited values of masses and radii.
\subsection{String lying on two extended and two compact dimensions.}
In this section we will consider a string lying on both the
compact dimensions of the space-time background
$\field{R}^{1,7}\times S^1\times S^1$, with say $X^1$ on one $S^1$ and
$X^3$ on the other $S^1$ and $X^{2,4}$ on $\field{R}^7$. Here we
expect a substantial contribution of the winding modes to the decay,
since the string can wind with both chiralities.

This has been the most difficult case to be studied: the number of
computer processes needed to calculate the imaginary part
of the amplitude (formula (\ref{imaginarypartsu2})) 
is very high. Furthermore increasing the values of the masses implies
working with very large numbers. Consequently our analysis
has been limited to values for the masses of:
 \beq
 M^2=19,29,39,59
 \eeq
and to a restricted range of radii (in the case of $M^2=59$ only a
few data could be obtained).

As we said, in this case the decay
through winding modes is possible, and for small radii (compared
to the string length) we expect it is the favorite decay.
Table (\ref{rateres2}) proves this claim for all the
mass tested: for $R<M$ the decay channels without windings give a
small fraction of the total rate, for $R\geq M$, on the contrary,
their contribution dominates.

\vspace{0.5cm}
\begin{figure}[h!]
\label{Tsu2} 
\vskip -0.5cm \hskip -1cm
\centerline{\epsfig{figure=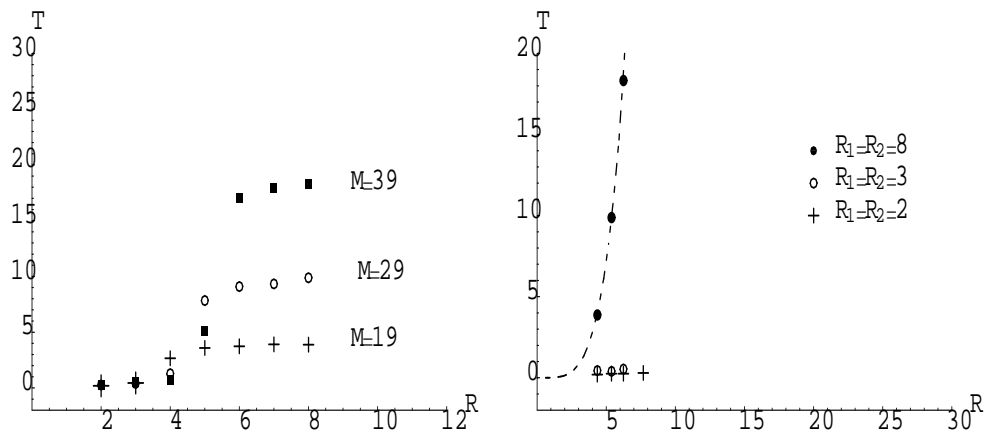,height=6truecm, width=14truecm}}
\caption[$R$ and $M$ dependence for the lifetime 
(string lying on two compact dimensions).]{\footnotesize
Case II: $R$ and $M$ dependence for the lifetime.The lines are fits to
the data.}
\end{figure}

\subsubsection{$R$ dependence.}
The power law (\ref{lifetimeexpected}) is not correct for radii
$R<M$: the decay rate is much more rapid for very small radii due
to a dominant contribution from channels of decay in winding modes, but
this contribution decreases rapidly to a negligible fraction of
the total rate of decay when increasing the radii.

We have reported in figure \ref{Tsu2}a the lifetime at fixed $M$
for various $R$.

For radii $R>M$, instead, as expected, the lifetime is flat with
respect to the radii.

\subsubsection{$M$ dependence.}
Due to the dominant contributions of the decays through winding modes, 
the expected power law (\ref{lifetimeexpected}) is not
correct for $R<M$. The lifetime is almost flat or slightly increases
for higher masses, and it does not exhibit a power law dependence on
mass, as can be seen in figure \ref{Tsu2}b.

For $R>M$, instead, the mass dependence of the lifetime tends to
agreement with 
(\ref{lifetimeexpected}), although we could test it only for radii
just above the threshold and for squared masses $M^2=19,\, 29,\, 39$: 

\begin{table}[htbp]
 \centering
 \begin{tabular}{||c|l|c||}
 \multicolumn{3}{|c|}{Range of masses: $4.36<M<6.24$}\\
 \hline
 Radii & $\beta $ & Expected value   \\
 \hline
 7 & 2.11488 & 2.5   \\
 8 & 2.16404 & 2.5     \\
 \end{tabular}
 \parbox{5in}{\caption[$M$ dependence for different radii 
 (string lying on two compact dimensions).]{
 $M$ dependence for different radii: $\T \sim M^{2*\beta}$. 
 }}\label{Mdependencesu2}  
 \end{table}

\subsection{Comparison of the various cases with each other.}

In this section we will discuss the results found for the decay rates
comparing the likelihood for the string to decay in the different
cases 0, I, II, defined in section \ref{generalresults}.
In all the above cases two out of the nine space dimensions are
compactified on circles with the same radius $R_1=R_2=R$.
 
We would expect that if the string lies on the compactified
dimensions and therefore has the possibility to wrap on them (if
the radii of the compact dimensions are smaller than the string length), the
decay should be enhanced, unless the Virasoro constraints make it
difficult. For larger radii, instead, there shouldn't
be any appreciable difference in the decay rates, since for all the
string configuration they should tend to the flat space value.

We have computed the ratio between the values for the decay rates in 
the various cases at
corresponding masses and radii, see table (\ref{decaycomparison}).

The results show that for radii $R>M$ the decay rates tend to the same
value (the flat space-time one). Let us analyze the behaviour for small
radii $R<M$.

 The rate for case II is the largest
one, with respect to both configurations I and 0. Furthermore
the ratio 
increases for higher masses at
fixed radii.  This is
expected, since the decay rates for the string in configuration I or 0
decreases with a power law as $\mathcal{R}\sim
M^{-1.5}$ , whereas the one for the string lying on two
compact dimensions decreases in a less marked way (see figure
(\ref{Tsu2}) and tables (\ref{lftmres}-\ref{rateres2}),
the rate is the inverse of the lifetime).

On the contrary the ratio between the decay rate for cases I and 0
shows, that in the first case (I), the string decays less. This can
be accounted by the fact that the string constraints $\dot{X}+X'=0, \,
\dot{X}\cdot X'=0$ are less easily satisfied in the more asymmetric
configuration of the string lying on one compact dimension only.

\begin{table}[htbp]
 \centering
 \begin{tabular}{||c|c|c|c||}
 \hline
 \multicolumn{4}{|c|}{Mass $M=4.36$}\\
 \hline
 Radii & $II/0$ & $I/0$ & $II/I$   \\
 \hline
 2 & 4.23898 & 0.732018   & 5.79082 \\
 3 & 4.04003 & 0.811177   & 4.98046   \\
 4 & 1.05918 & 0.914973   & 1.15761  \\
 5 & 0.974573  & 0.972415 & 1.00222   \\
 6 & 1.01443 & 1.00094    & 1.01348  \\
 7 & 0.999821 & 1.00312   & 0.996712  \\
 8 & 1.02703 & 1.01037    & 1.01649  \\
 \hline
 \multicolumn{4}{c}{}\\
 \hline
 \multicolumn{4}{|c|}{Mass $M=6.24$}\\
 \hline
 Radii & $II/0$ & $I/0$ & $II/I$   \\
 \hline
 2 & 8.91015 & 0.641493   & 13.8897 \\
 3 & 9.42839 & 0.685099   & 13.7621   \\
 4 & 13.0199 & 0.777693   & 16.7417  \\
 5 & 2.47845  & 0.869407  & 2.85074  \\
 6 & 0.911707 & 0.943306    & 0.966502  \\
 7 & 0.975422 & 0.975233   & 1.00019  \\
 8 & 1.01423 & 0.994766    & 1.01956  \\
 \hline
 \multicolumn{4}{c}{}\\
 \hline
 \multicolumn{4}{|c|}{Mass $M=7.68$}\\
 \hline
 Radii & $II/0$ & $I/0$ & $II/I$   \\
 \hline
 2 & 13.7567 & 0.624082   & 22.0431 \\
 4 & 21.2468 & 0.709832   & 18.5974  \\
 \hline
 \end{tabular}
 \parbox{5in}{\caption[Decay rates comparison.]{
 Decay rates comparison for some fixed mass and radii. 
 
 II means string lying on two compact dimensions with opposite chiralities.

 I means string lying on one compact dimension.

 0 means string lying on extended dimensions only.
 }\label{decaycomparison}}  
 \end{table}

\section{Summary of the results and conclusions.} \label{conclusion}

We have studied the decay of a particular closed string state of type
II string theory that in a previous work (\cite{CIR}) was shown to be
long-lived in ten uncompactified flat dimensions. It classically
corresponds to the string configuration of equation (\ref{solution}).
 
We have considered a space-time $\field{R}^{1,7}\times
T^2$ and chosen three string configuration:
 \begin{itemize}
 \item[-]0: string lying on extended dimensions only;
 \item[-]I: string lying on three extended and one compact dimension; 
 \item[-]II: string lying on two extended and two compact dimensions,
 with opposite chirality.
 \end{itemize} 
Throughout all the paper we have considered $R_1=R_2=R$, but we have
also data (not shown) for $R_1\neq R_2$.

In case 0, the string decay is entirely given by the emission of a
massless plus a massive final state. The difference between this and
the ten flat uncompactified dimensions case lies only in the reduced
phase-space factor when the radii of compactification are small.

The string in configuration I decays in one massless and one massive
state for $R<<M$. For $R\sim M$, the other dominant decay channel is
through one Kaluza-Klein plus one massive final state. This channel
becomes the most important for $R>M$, where the total decay rate
flattens as expected and the Kaluza-Klein modes play the role of
momentum components. The winding modes do not seem to play a
significant role at any radius.

Finally, in case II the string shows the most rapid and intense decay
for small radii. The winding modes channels 
dominate. For $R<M$ the lifetime slightly increases by increasing the
parent mass (at fixed radii) and by increasing the  radii $R<M$ (at
fixed mass). For
$R>M$ we recover the uncompactified dimensions result.

In all the cases the threshold separating the behaviour at large radii
(lifetimes $\mathcal{T}\sim g_s M^5$ and no dependence on $R$, as in
the uncompactified scenario) from the behaviour at small radii
($\mathcal{T}$ increasing more slowly) is found to be $R\sim M$.

The above results confirm the general expectations on the effect of
compactification
on the decay properties of the string state we have considered. Although we
have examined in detail the case in which two of the nine space dimensions
of the theory are compact, the decay pattern can be clearly extrapolated to
the case of compactification of additional dimensions. 

Our conclusion is that the string configuration we have studied is generally
long lived. The lifetime increases with the mass, at least when
its size is smaller than the radius of the dimensions it lies on,
and even for larger sizes, if it lies in part on the uncompact space 
and it does not wind with opposite chiralities on the compact dimensions.
The dominant decay channel is the soft emission of massless states. 

\section{Acknowledgments}
Partial support by the EEC Network contract HPRN-CT-2000-00131 and by
the Italian MIUR program ``Teoria dei Campi, Superstringhe e
Gravit\`a'' is acknowledged.


\appendix
\section{Appendix A: Computation of ${\rm Im}(\Delta M^2)$}\label{appendiximm}

In these appendices we explain in details the computations needed to
obtain the imaginary part of amplitude (\ref{vertexamplitude}) shown
in section (\ref{imaginarypart}).

We have always taken two space-time coordinates as compactified on
a torus, while all the others were not. It means that we identify
points of the two compactified dimensions under the action of the
group of translation generated by the element
 \beq
  v=2\pi R
 \eeq
so that the bosonic coordinates transform as
 \beq
 X^\mu=X^\mu +m v,\quad m\,\epsilon\, \field{Z},
 \eeq
the fermionic
 \beq
 \psi^\mu=\psi^\mu
 \eeq
and the resulting space-time is:
 \[\field{R}^8\times T^2.\]

This implies that the fermionic propagator is unchanged and still
obeys the Riemann identity, the bosonic one, instead, is different
depending on if the string lies or not on the compactified
coordinates.

This propagator can be computed trough path integral formalism.


\subsection{String lying on extended dimensions only.} \label{appendixsu0}

The propagator for the string coordinates in this case is the same as
 in the uncompactified scenario:

 \be \label{flatpropagator}
 \langle \p\bar Z_i(z)\p Z_j(0)\rangle =\delta_{ij}\ 2\left(
 \p^2  \log \theta_1(z|\tau )+{\pi\ov\tau_2}\right) \ ,\nonumber\\
 \langle \bar\p\bar Z_i(z)\bar\p Z_j(0)\rangle = \delta_{ij}\
 2\left( \bar\p^2 \log \bar\theta_1(z|\tau )+{\pi\ov\tau_2}\right)\,\nonumber\\
 \langle \bar\p\bar Z_i(z)\p Z_j(0)\rangle =
 \langle \p\bar  Z_i(z)\bar\p Z_j(0) \rangle =-\delta_{ij}2\ {\pi\ov\tau_2}\ ,\nonumber\\
 \langle e^{-ipX(z)}e^{ipX(0)} \rangle =e^{-4N{\pi y^2\ov\tau_2}}\
 \left| {\theta_1 (z|\tau )\ov\theta_1^{'} (0|\tau)}\right| ^{4N}
 \ee
 (here $y={\rm Im} (z)$).

What changes is the integral over the zero modes (here not only
momenta, but also winding modes) giving us a final measure of
integration for $\tau$
 \beq \label{measureofintegration}
 \int \frac{d^2\tau}{\tau_2^4}\prod_{i=1}^2\left({1\ov\sqrt{\pi\tau_2}}
 \sum_{l_i,w_i}e^{-\frac{\pi R^2}{4\tau_2}|l_i-m_i\tau|^2}\right)
 \eeq
where $l_i,\,m_i$ are winding modes.

The correlator in (\ref{vertexamplitude}) is:
 \be \label{correlator}
 \langle \bar V_{k,k}(z)V_{k,k}(0)\rangle & = & \\
 & = & {N^2\ov 4^{N-1}((N-1)!)^2}\langle \langle e^{-ipX(z)}e^{ipX(0)} \rangle 
       \nonumber \\
 & \times & \psi_{\bar z_1}(z)\p\psi_{\bar z_1}(z)
       \td\psi_{\bar z_2}(z)\bp\td\psi_{\bar z_2}(z)
       \psi_{z_1}(0)\p\psi_{z_1}(0)\td\psi_{z_2}(0)\bp\td\psi_{z_2}(0)\rangle 
       \nonumber \\
 & \times  & \langle (\p \bar Z_1(z))^{N-1}(\p Z_1(0))^{N-1}
       (\bp \bar Z_2(z))^{N-1}(\bp Z_2(0))^{N-1}\rangle
        \nonumber 
 \ee
giving, from (\ref{flatpropagator}) and
 \beq
 \langle (\p \bar Z_1)^{N-1}(\p Z_1)^{N-1}\rangle=(N-1)!\delta_{ij}\ 2\left(
 \p^2  \log \theta_1(z|\tau )+{\pi\ov\tau_2}\right),
 \eeq
the result:
 \beq
 \langle \bar V_{k,k}(z)V_{k,k}(0)\rangle  =(N)^2 \langle e^{-ipX(z)}e^{ipX(0)} \rangle
       \left| \p^2 \log \theta_1(z|\tau)+{\pi\ov\tau_2} \right| ^{2(N-1)}
 \eeq
as in \cite{CIR}.

 We expand then the holomorfic
and antiholomorfic factors in binomials, obtaining:
 \be
  && \langle \bar V_{k,k}(z)V_{k,k}(0)\rangle =
      (N)^2 \langle e^{-ipX(z)}e^{ipX(0)} \rangle ({\pi \ov \tau_2})^{2(N-1)}\\
  && \times\sum_{m_1,m_2} \big( {\pi\ov\tau_2} \big) ^{-m_1-m_2}
      \mbox{\small{$\frac{((N-1)!)^2}{m_1!m_2!(N-1-m_1)!(N-1-m_2)!}$}}\big( \p^2 \log(\theta_1(z|\tau ) \big)^{m_1}
     \big(\bar\p^2\log(\theta_1(z|\tau ) \big)^{m_2} \ . \nonumber
 \ee

We perform a Poisson re-summation on the sum over $l_i$ in 
(\ref{measureofintegration}) so that we
obtain the final amplitude (\ref{amplitudeextendedstring}).

\subsection{String lying on one compactified dimensions.}  \label{appendixsu1}

In this case, coordinate $X^1$ is compactified (together with another
coordinate, where the string does not lies).

According to (\ref{complexcoord}) for the correlator $\langle \bar
V V\rangle$ we need in particular the part
 \be \label{pipropagator}
  &\mbox{}& \langle \p X^1(z_1)\p X^1(z_2) \rangle  =  \\
  &\mbox{}& \sum_{m,l}\int_{\substack{X^1(z+1)=X^1(z)+2\pi Rm \\
                                      X^1(z+\tau)=X^1(z)+2\pi Rl}}
    \mathcal{D}X^1
      e^{-S[X^1,\tau]}\p X^1(z_1)\p X^1(z_2) \nonumber
 \ee
To compute this path-integral we can split the coordinate $X^1(z)$
in a classical part, obeying the boundary conditions, and a
periodical quantum part
 \beq
 X^1(z)=X^1_{cl}(z)+X^1_{qu}(z)
 \eeq
where
 \beq
 X^1_{cl}(z)=\frac{\pi R}{i\tau_2}(l-m\bt)z-\frac{\pi R}{i\tau_2}(l-m\tau)\bar z
 \eeq
In terms of the complex coordinates (\ref{splitcoord}), we write
 \be \label{splitcoord}
 \begin{cases}
 \p Z_1(z) & =  \p Z_1^0(z)+{\alpha\ov\sqrt 2} \\
 \p \bar Z_1(z) & =  \p \bar Z_1^0(z)+{\alpha\ov\sqrt 2} \\
 \end{cases}
 \ee
where
 \beq
 \alpha={\pi R_1\ov i\tau_2}(l-m\bt)
 \eeq
and $\p Z_1^0(z)$ is the oscillator part of the string coordinate
expansion.

The correlator is again given by (\ref{correlator}), but now
equations (\ref{splitcoord}) lead to:
 \be
  \langle (\p \bar Z_1)^{N-1} (\p Z_1)^{N-1}\rangle & = & \langle \left(\p \bar Z^0_1+{\alpha\ov\sqrt 2}\right)^{N-1}
              \left(\p Z^0_1+{\alpha\ov\sqrt
              2}\right)^{N-1}\rangle= \\
  & = & \sum_{s_1,s_2=0}^{N-1}
         \mbox{\tiny{$\left(\begin{array}{c} N-1\\s_1 \end{array}\right)
         \left(\begin{array}{c} N-1\\s_2 \end{array}\right)$}}\langle\p\bar
         Z_1^0(z)^{N-1-s_1}\p Z_1^0(0)^{N-1-s_2}\rangle \nonumber \\
   & \times & \mbox{\small{$\left({\alpha\ov \sqrt 2}\right)^{s_1}
         \left({\alpha\ov \sqrt 2}\right)^{s_2}$}}\nonumber
 \ee
so that the result is:
 \be
 \langle \bar V(z)V(0)\rangle & = & \\
   & = & {N^2\ov 4^{N-1}((N-1)!)^2}\sum_{s_1,s_2=0}^{N-1}
         \mbox{\tiny{$\left(\begin{array}{c} N-1\\s_1 \end{array}\right)
         \left(\begin{array}{c} N-1\\s_2 \end{array}\right)$}} \nonumber \\
   & \times  & \langle\p\bar
         Z_1^0(z)^{N-1-s_1}\p Z_1^0(0)^{N-1-s_2}\rangle \left({\alpha\ov \sqrt 2}\right)^{s_1}
         \left({\alpha\ov \sqrt 2}\right)^{s_2} \nonumber \\
   & \times  & \langle\bp\bar Z_2^0(z)^{N-1}\bp Z_2^0(0)^{N-1}\rangle
         e^{-4N{\pi y^2\ov\tau_2}}
         \left| {\theta_1 (z|\tau )\ov\theta_1^{'} (0|\tau)}\right| ^{4N}
 \nonumber
 \ee
The non-vanishing part of the amplitude is the one with
$s_1=s_2=s$. Using, then, formulas (\ref{flatpropagator}) for the
correlator of the oscillator part of the string coordinate and 
(\ref{measureofintegration}) for the measure of integration, we
arrive to the result:
 \be \label{amp1conpstraarr}
 \Delta M^2 & = & c' \ g_s^2\ \int {d^2\tau\ov\tau_2^4}
              \int d^2z\sum_{l_{1,2},w_{1,2}}\sum_{s=0}^{N-1}\sum_{m_1=0}^{N-1-s}\sum_{m_2=0}^{N-1}
              \mbox{\small{${N^2((N-1)!)^2\ov (s!)^2m_1!m_2!(N-1-s-m_1)!(N-1-m_2)}$}} \\
         & \times & \left({\pi\ov\tau_2}\right)^{2N-2}\left(-{\pi R_1^2\ov 4\tau_2}(l_1-w_1\bar\tau)^2\right)^s
              \left({\tau_2\ov\pi}\p^2\log\theta_1(z|\tau)\right)^{m_1}
              \left({\tau_2\ov\pi}\bp^2\log\bar\theta_1(\bar z|\bar\tau)\right)^{m_2}  \nonumber \\
         & \times & e^{-4N{\pi y^2\ov\tau_2}}
              \left| {\theta_1 (z|\tau )\ov\theta_1^{'}(0|\tau)}\right|^{4N}\frac{1}{\pi\tau_2}
              e^{-{\pi\ov 4\tau_2}R_1^2|l_1-w_1\tau|^2-{\pi\ov 4\tau_2}R_2^2|l_2-w_2\tau|^2}
            \nonumber
 \ee
Now consider:
 \beq \label{rightmovingcompcontr}
 A_R=\sum_{l_1}\sum_s\left({\pi R_1^2\ov 4\tau_2}(l_1-w_1\bar\tau)^2\right)^s e^{-{\pi\ov
 4\tau_2}R_1^2|l_1-w_1\tau|^2},
 \eeq
where the subscript $R$ refers to the fact the this concerns the
right moving part of the string only.

Call
 \beq
 a={\pi R_1^2\ov 4\tau_2}.
 \eeq
We can write:
 \be
 A_R & = & \sum_{l_1}\sum_s\left(a(l_1-w_1\bar\tau)^2\right)^s
          e^{-a|l_1-w_1\tau|^2} \\
   & = & \left.\sum_{l_1}\sum_s(-\p_b)^s
          e^{-a|l_1-w_1\tau|^2-ba(l_1-w_1\bar\tau)^2}\right|_{b=0}.
 \nonumber
 \ee
Now perform a Poisson re-summation on $l_1$:
 \be
 A_R & = & \left.\sum_{l_1}\sum_s(-\p_b)^s
          e^{-a|l_1-w_1\tau|^2-ba(l_1-w_1\bar\tau)^2}\right|_{b=0} \\
   & = & \left.\sum_{n_1}\sum_s\sqrt{{\pi\ov a}}(-\p_b)^s
          \sqrt{{1\ov 1+b}}
          e^{-{\pi^2\ov
          a(1+b)}\left(n_1+{aw_1\tau_2\ov\pi}\right)^2+2i\pi\tau_1n_1w_1+2\pi w_1\tau_2n_1}\right|_{b=0}.
          \nonumber
 \ee
We have got to compute
 \beq
 P=\sum_s(-\p_b)^s
     \sqrt{{1\ov 1+b}}
     e^{-{\pi^2\ov
     a(1+b)}\left(n_1+{aw_1\tau_2\ov\pi}\right)^2}.
 \eeq
Calling therefore $y={1\ov 1+b}$ it can be shown that
 \beq
 \sum_s(-\p_b)^s=\sum_s(y^2\p_y)^s=\sum_s\sum_{l=1}^s c_s^ly^{l+s}\p_y^l
 \eeq
where the coefficients $c_s^l$ are defined recursively as
 \be \label{cslcoefficients}
 \begin{cases}
 c_s^l &=  c_{s-1}^l(l+s-1)+c_{s-1}^{l-1} \\
 c_s^l &=  0, \, s<l \nonumber \\
 c_s^1 &=  s! \, s\neq 0 \nonumber \\
 c_s^0 &=  0,\, s\neq 0 \nonumber \\
 c_0^0 &=  1 \nonumber \\
 \end{cases}
 \ee
 so that
 \beq
 P=\sum_s\sum_{l=0}^s c_s^l\sum_{r=0}^l
  \mbox{\small{$\left(\begin{array}{c} l\\r \end{array}\right)$}}
  {\Gamma(3/2)\ov\Gamma(3/2-l+r)}\, e^{-{4\pi\tau_2\ov
     R_1^2}\left(n_1+{w_1R_1^2\ov 4}\right)^2}\left(-{4\pi\tau_2 \ov R_1^2}\left(n_1+{w_1R_1^2\ov 4}\right)^2\right)^r.
 \eeq
Inserting this result in (\ref{amp1conpstraarr}), performing a
Poisson re-summation also on $l_2$ and expanding again the relevant
terms as in (\ref{pqexpansion}), we get an imaginary part for the mass
 shift as shown in equation (\ref{imaginarypartsu1}).

\subsection{String lying on two compactified dimensions.}  \label{appendixsu2}

In this case we have compactified both coordinates $X^1$ and
 $X^3$. For the Right-moving part of the
string, the contribute to the amplitude is the same as the Right
moving part discussed in the previous section. For what concerns
instead the Left-moving sector, the computation is analogous,
substituting
 \beq
  A_L=\sum_{l_2}\sum_s\left({\pi R_2^2\ov 4\tau_2}(l_2-w_2\tau)^2\right)^s e^{-{\pi\ov
    4\tau_2}R_2^2|l_2-w_2\tau|^2}
 \eeq
to (\ref{rightmovingcompcontr}).

 With the same steps as in the previous computation, eventually we obtain:
 \be
 A_L & = & \sum_{n_2}\sum_s\sum_{l=0}^s c_s^l
          \mbox{\small{$\left(\begin{array}{c} l\\r \end{array}\right)$}}
          {\Gamma(3/2)\ov\Gamma(3/2-l+r)}\,\\
   & \times &  e^{-{4\pi\tau_2\ov R_2^2}\left(n_2-{w_2R_2^2\ov 4}\right)^2+2i\pi\tau_1w_2n_2-2\pi\tau_2w_2n_2}
                \left(-{4\pi\tau_2 \ov R_2^2}\left(n_2-{w_2R_2^2\ov 4}\right)^2\right)^r.
 \ee

The imaginary part of the amplitude is, then, shown in equation 
(\ref{imaginarypartsu2}).

\section{Appendix B: Tables}\label{appendixtables}

We report here a selection of the data.
\vspace{2cm}

\enlargethispage{\baselineskip}
\begin{table}[!hp]
\begin{tabular}{||c|c|c|c|c||}
\hline
$M$ & $R1=R2=R$ & $ dw=0$ & $ dw=1$ 
       \\ \hline
\hline
$\sqrt{19}=4.36$ & $2$ & $1.2172\,(1.2167)$ 
                 & $0.89104\,(0.81904)$ 
\\ 
\hline
%
$\sqrt{19}=4.36$ & $4$ & $0.35470\,(0.35465)$ 
       & $0.32454\,(0.32345)$ 
\\ 
\hline
$\sqrt{19}=4.36$ & $8$     & $0.25114\,(0.25109)$     &  $0.25374\,(0.25369)$   
\\ 
\hline
$\sqrt{19}=4.36$ & $16$     & $0.24990(0.24985)$  & $0.24963(0.24959)$     
\\ 
\hline
\hline
$\sqrt{39}=6.24$ & $2$ & $0.44313\,(0.44313)$ & $0.28427\,(0.28057)$ 
\\ 
\hline
%
$\sqrt{39}=6.24$ & $4$ & $0.11406\,(0.11406)$ 
            & $0.088708\,(0.088681)$ 
\\ 
\hline

$\sqrt{39}=6.24$ & $8$ & $0.053780\,(0.053779)$ 
          & $0.05350\,(0.053498)$ 
\\ \hline
$\sqrt{39}=6,24$ & $16$     & $0.050159\,(0.050159)$ 
                & $0.050175\,(0.050175)$ 
\\  
\hline
\hline
$\sqrt{59}=7.68$ & $2$ & $0.24404\,(0.24404)$ 
           & $0.15230\,(0.15177)$ 
\\ \hline
$\sqrt{59}=7.68$ & $4$ & $0.061474\,(0.061474)$ 
              & $0.043636\,(0.043635)$ 
\\ \hline
$\sqrt{59}=7.68$ & $8$ & $0.022650\,(0.022650)$ 
      & $0.0218738\,(0.2187)$ 
\\
\hline
$\sqrt{59}=7.68$ & $16$     & $0.01919(0.019187)$   & $(0.2496)$ 
\\ 
\hline
\hline
\end{tabular}
\caption[Decay Rates (string lying on one or no compact dimensions).]
 {\small{Decay Rates for two compact dimensions
 ($dw$=number of compact dimensions where the string
 lies):

 - numbers not enclosed in brackets represent
 the total  decay rate, 
 
 - numbers in round brackets ($\dotsc$) represent the
 contribution of channels where one of the two decay
 products is a {\it pure Kaluza-Klein} mode. Actually, for small radii the
 decay in non-zero Kaluza-Klein modes 
 represents only less than 0.5\% of the
 total decay for large masses in the case $dc=2,\, dw=0$ and between
 the 0.6\% and 13\% for the case $dc=2,\, dw=1$.


 

 }}\label{rateres01}

\vspace{0.5cm}
\begin{tabular}{||c|c|c||}
\hline
$M$ & $R1=R2=R$ &        $dw=2$  \\ \hline
\hline
$\sqrt{19}=4.36$ & $2$ & $5.1598\,\{0.66702\}$ \\ 
\hline
%
$\sqrt{19}=4.36$ & $4$ &  $0.37569\,\{0.30917\}$ \\ 
\hline
$\sqrt{19}=4.36$ & $8$ &       $0.25793\,\{ 0.25793\}$\\ 

\hline
\hline
$\sqrt{39}=6.24$ & $2$ &        $3.9484\,\{0.22226\}$ \\ 
\hline
%
$\sqrt{39}=6.24$ & $4$ & $1.4851\,\{0.077970\}$ \\ 
\hline

$\sqrt{39}=6.24$ & $8$ &            $0.054544\,\{0.054544\}$\\
\hline
\hline
$\sqrt{59}=7.68$ & $2$ & $3.3572\,\{0.11935\}$ \\ \hline
$\sqrt{59}=7.68$ & $4$ & $1.3061$ \\
\hline
\hline
\end{tabular}
\caption[Decay Rates (sting lying on two compact dimensions).]
 {\small{Decay Rates for two compact dimensions 
 (here $dw$=number of compactified dimensions where the string lies=2):

 - numbers not enclosed in brackets represent
 the total  decay rate, 
 
- numbers in curly brackets \{$\dotsc$\} represent the
 contribution of channels with {\it zero winding modes}.

 }}\label{rateres2}
\end{table}

\vskip0.5truecm

\begin{table}
\begin{center}
\begin{tabular}{||c|c|c|c|c||}
\hline
$M$ & $R1=R2=R$ & $dw=0$ & $dw=1$ & $dw=2$  \\ \hline
\hline
$\sqrt{19}=4.36$ & $2$ & $0.82153$ & $1.1223$ & $0.19380$ \\ \hline
$\sqrt{19}=4.36$ & $4$ & $2.8193$ & $3.0813$ & $2.6617$ \\ \hline
$\sqrt{19}=4.36$ & $8$     & $3.9819$     & $3.9410$     & $3.8771$\\ \hline
$\sqrt{19}=4.36$ & $16$     & $4.0017$     & $4.0058$     & $~$\\ \hline
\hline
$\sqrt{39}=6.24$ & $2$ & $2.2566$ & $3.5178$ & $0.25327$ \\ \hline
%
$\sqrt{39}=6.24$ & $4$ & $8.7669$ & $11.2729$ & $0.67335$ \\ \hline
$\sqrt{39}=6.24$ & $8$ & $18.594$ & $18.6924$ & $18.334$\\ \hline
$\sqrt{39}=6.24$ & $16$     & $19.936$     & $19.930$     & $~$\\ \hline
\hline
$\sqrt{59}=6.24$ & $2$ & $4.0977$ & $6.5660$ & $0.29787$ \\ \hline
$\sqrt{59}=6.24$ & $4$ & $16.267$ & $22.9169$ & $0.76563$ \\ \hline
$\sqrt{59}=6.24$ & $8$ & $44.150$ & $45.7167$ & $~$\\ \hline
$\sqrt{59}=6.24$ & $16$     & $52.118$     & $~$     & $~$\\ \hline
\hline
\end{tabular}\end{center}
\caption[Values of the lifetime]{Values of the lifetime for two
  compact dimensions. ($dw$=number of compact dimensions where the string
  lies)}\label{lftmres}
\end{table}


\end{document}